\documentclass[letterpaper,twocolumn,10pt]{article}
\usepackage{usenix2019_SOUPS}

\usepackage{tikz}
\usepackage{amsmath}

\usepackage{filecontents}

\usepackage{graphics} 
\usepackage{comment}

\usepackage{booktabs} 

\usepackage{multirow}

\usepackage[ruled]{algorithm2e} 

\usepackage{mathtools}
\DeclarePairedDelimiter\abs{\lvert}{\rvert}%
\DeclarePairedDelimiter\norm{\lVert}{\rVert}%

\usepackage{amsthm}
\theoremstyle{definition}
\newtheorem{definition}{Definition}[section]

\newtheorem{property}{Property}[section]

\usepackage{color, colortbl}
\definecolor{mary1}{rgb}{0.969,0.816,0.541}
\definecolor{mark2}{rgb}{0.890,0.941,0.608}
\definecolor{mary34}{rgb}{0.529,0.714,0.655}

\usepackage{amsfonts}

\begin{document}

\date{}

\title{\Large \bf Increasing Transparent and Accountable Use of Data by Quantifying the Actual Privacy Risk in Interactive Record Linkage}

\def\plainauthor{Author name(s) for PDF metadata. Don't forget to anonymize for submission!}

\author{
{\rm Qinbo Li ${^1}$}\\
\and
{\rm Adam G. D'Souza ${^2}$}\\
\and
{\rm Cason Schmit ${^1}$}\\
\and
{\rm Hye-Chung Kum ${^1}$}\\
\and
${^1}$Texas A\&M University, College Station, TX \hspace{0.5cm} ${^2}$University of Calgary, Calgary, AB, Canada
} 

\maketitle

\begin{abstract}

Record linkage refers to the task of integrating data from two or more databases without a common identifier. MINDFIRL (MInimum Necessary Disclosure For Interactive Record Linkage) is a software system that demonstrates the tradeoff between utility and privacy in interactive record linkage. Due to the need to access personally identifiable information (PII) to accurately assess whether different records refer to the same person in heterogeneous databases, privacy is a major concern in interactive record linkage.
MINDFIRL supports interactive record linkage while minimizing the privacy risk by (1) using pseudonyms to separate the identifying information from the sensitive information, (2) dynamically disclosing only the minimum necessary information incrementally, as needed on-demand at the point of decision, and (3) quantifies the risk due to the needed information disclosure to support transparency, the reasoning, communication, and decisions on the privacy and utility trade off. In this paper we present an overview of the MINDFIRL system and the k-Anonymized Privacy Risk (KAPR) score used to measure the privacy risk based on the disclosed information. We prove that KAPR score is a norm meeting all the desirable properties for a risk score for interactive record linkage.
\end{abstract}

\section{Introduction}
\label{sec:Introduction}
As data fuel the knowledge economy, appropriate collection and use of personal data for better decisions is increasing rapidly for all purposes including business, government, and research.
Yet, there are also increased concerns about individual privacy leading to the need for improved methods in private data analysis. Research on information privacy has shown that, mathematically, private data analysis requires a tradeoff between privacy and utility because all use of personal data leads to some privacy loss while providing some utility in terms of data analysis.
Yet, less is known about how to make practical decisions on the privacy-utility tradeoff in real applications that are often dependent on the specifics of the data task and the data being used. 

For record linkage and integration tasks that lack a common identifier, fully-automated systems have limitations because of the many data problems in real data such as duplicate records, non-unique names, data errors in unique identifiers (e.g., SSN), non-stable attributes (e.g., change in last name), and valid similar identifiers for different people (e.g., twins, Jr/Sr).
Differentiating between whether two records are duplicates of the same person or two near-identical records (e.g., as is often the case for twins) often requires additional information gathering and human judgment.
Most systems that are used in practice are semi-automated interactive record-linkage systems where human judgment is used to clean, standardize, and manually resolve uncertain matches.
Thus, privacy becomes a major concern when humans are accessing personally identifiable information (PII) for accurate record linkage.
In this demonstration, we present a system and method focusing on privacy-utility tradeoff decisions in the task of record linkage for integrating heterogeneous person-level databases without a common identifier.

\section{Related Work}
\label{sec:Background}
Classic results in information privacy show that one cannot answer more than an adversarially chosen set of $n(log n)^2 $
queries over a database of $n$ bits such that each query has $o(\sqrt{n}$) errors without the adversary being able to reconstruct the original database \cite{Dinur2003}. 
This result poses a fundamental limit on private data analyses and motivates the need to think about private data analysis as a budget-constrained problem.
That is, each use of data leads to some privacy loss while providing some utility in terms of data analysis. 
This has led to a rich body of literature on differential privacy as a method that lets us concretely reason about  privacy-budgeted data analysis.
In particular, many differentially private algorithms have been proposed to answer low dimensional counting queries.
However, adoption of these methods in practice has been limited due in part to the wide variation in error rates, which are dependent on the properties of the input data.
 To facilitate adoption, Kotsogianis et al. have recently released PYTHIA, a software tool for learning the tradeoff between accuracy and privacy in simple count queries \cite{Kotsogiannis2017}. MINDFIRL is a similar software that supports learning the privacy and utility tradeoff in interactive record linkage and
provides optimal parameters to control fine grained access via incremental information disclosure.

 Private record linkage (PRL) is the most common approach to protecting privacy in record linkage. PRL allows the disclosure of the linked data to two parties while prohibiting the disclosure of unlinked data to the other party or any third party \cite{vatsalan2013}.
 In PRL, an accurate linkage function is usually assumed to be available, 
 which dictates which data is linked and thus can be shared.
 Thus, the basic goal of PRL is to securely compute a known linkage function on encrypted data.
 The main limitation of these methods is that, in practice, such linkage functions are rarely known.
 Furthermore, there is no way to evaluate the quality of the linkage results.
 Thus, adoption of these methods in practice have been limited.

To overcome these issues, in our prior work, we proposed the Privacy Preserving Interactive Record Linkage (PPIRL) computational framework as a human-machine hybrid system with specific utility and privacy requirements that allows human interaction while preserving information privacy during the record linkage process \cite{kum2014}.
The privacy objective of PPIRL is to protect against sensitive attribute disclosure (e.g., cancer status) and minimize identity disclosure while its utility objective is to generate the optimal matching function possible by allowing for humans to fine tune the automated results by interacting with the data.
Such a computational framework allows for the reasoning of the privacy and utility tradeoff in record linkage.
MINDFIRL is a software system that meets all requirements of the PPIRL framework.

The dynamic visual interface for MINDFIRL was designed based on a prior large user study (N=104) on information requirements of interactive record linkage \cite{ragan2018}.
Using five different static interfaces with variations on information disclosed and markups on the discrepancies in the data, we were able to demonstrate the effectiveness of partial information disclosure for record linkage decisions.
We found that study participants who viewed only 30\% of data content had decision quality similar to those who had full 100\% access.
In addition, although quality of decision dropped significantly with too little information, the study demonstrated that with proper interface designs, many correct decisions can be made with even legally de-identified data that is fully masked (74.5\% accuracy with fully-masked data compared to 84.1\% with full access).

\section{Mindfirl}
\label{sec:Mindfirl}

The MINDFIRL software supports interactive record linkage via an on-demand information disclosure model and an interactive visual interface for linkage decisions.
The software supports two user roles:  manager and data worker.
The manager is the owner of the record linkage task (e.g., project investigator), while the workers are the people who actually work on the record linkage tasks (e.g., graduate students).
The manager can create record linkage tasks, import databases that need to be linked, set up a privacy budget using the systems's privacy risk settings, and dispatch the task to workers.
The worker can then work on the record linkage task within the privacy budget limit.

\begin{figure*}[h]
\includegraphics[width=\textwidth]{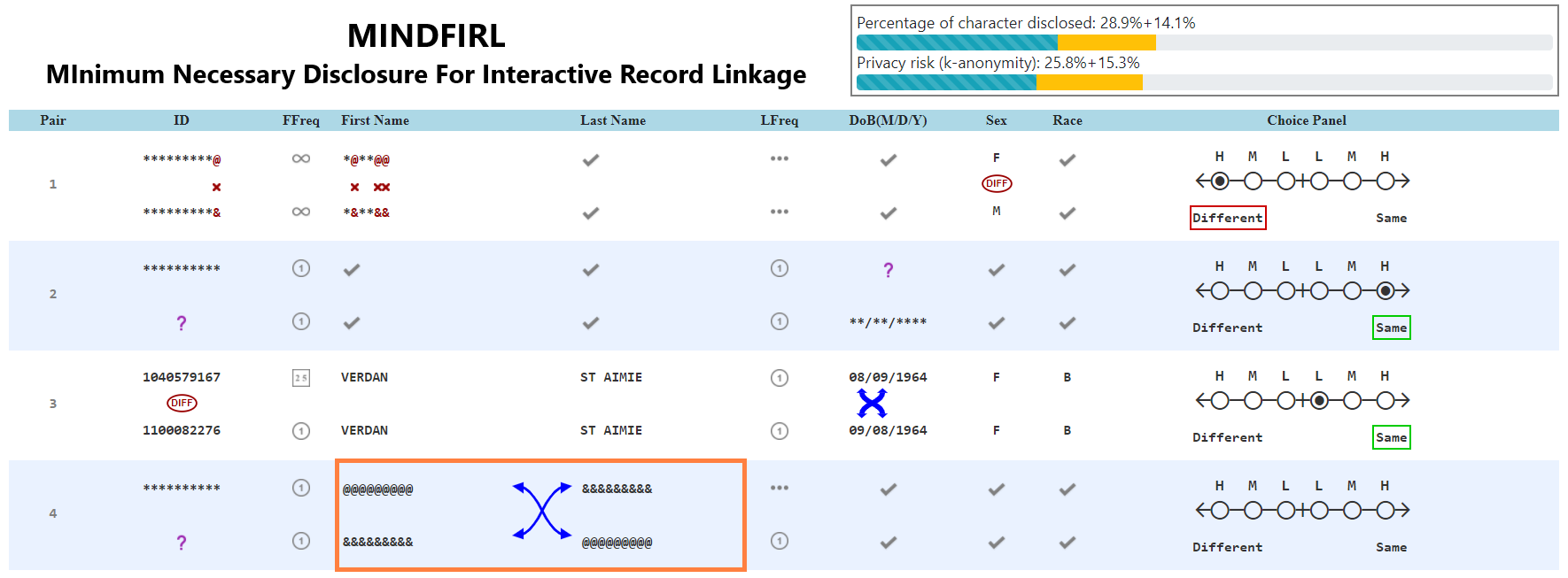}
\caption{The user interface for the record linkage tasks in MINDFIRL}
\label{fig:ui}
\end{figure*}

The record linkage problem-solving environment is shown in Figure \ref{fig:ui}.
The interface shows pairs of records that are potential matches but with some level of uncertainty.
The worker can incrementally reveal additional data details as needed to make better record linkage decisions.
Each time more information is opened, there is a cost associated with the information being revealed, and this cost is subtracted from the budget, as visually shown by meters on the top right of the interface.
During the interaction process, MINDFIRL pre-calculates the privacy risk for all the next potential states.
Therefore, the worker can check the cost of revealing a cell by hovering on the cell (e.g. orange box in last pair) and then decide whether or not to reveal the additional details.


\subsection{MINDFIRL Architecture}

MINDFIRL uses the client server model.
At any given time, the client software only has the information that has been revealed to the user and will need to request any additional details from the server when the user requests it.
This is to prevent users hacking into the client software to see more than is allowed.

An important property of the PPIRL computational framework is that the privacy requirement states that the software must guarantee no sensitive attribute disclosure (e.g., cancer status).
MINDFIRL meets this requirement by using the standard pseudonym mechanism to separate the sensitive attribute from the identifying attributes via a random number.
The random number is managed by the software to construct new linked data as needed without a person ever having to interact with the computer generated number.
The workers completing the record linkage task only interact with the identifying attributes, such as ID, name, gender, race, and are never shown the sensitive attributes.
Once the record linkage is done, MINDFIRL will provide the linked de-identified data for further analysis, but this time with no identifying information.

\subsection{On-Demand Incremental Disclosure}


Combining individual level information (e.g., summation) is the most common form of data manipulation to protect privacy when details in the data reveal too much personal information.
In record linkage tasks, workers are often given a pair of identifying records to determine if the two refer to the same real world person by taking into account the discrepancies in the two records.
Thus, summarizing the pair of records into information about the discrepancy of the pair has the dual affect of supporting good decision-making by distilling the raw data into more useful information as well as protecting privacy by hiding the individual level data.
Using this intuition, we designed different display modes and markups to support effective on-demand incremental information disclosures.

The system's \textit{data display mode} controls how the system displays an attribute.
Each attribute has three data display modes: \textit{masked disclosure}, \textit{partial disclosure}, and \textit{full disclosure}.
In \textit{masked disclosure} mode, all the characters are masked by symbols.
For \textit{partial disclosure} mode, some characters are masked while the other are shown based on the type of the attribute and the actual discrepancy of the value in the pair being reviewed for a linkage.
In \textit{full disclosure} mode, all characters are disclosed.
There are supplemental visual markups to aid the user to know what the discrepancies are to make a decision.
These markups can indicate issues such as missing values, differing values, identical values, swaps between values, and string-edit operations between values \cite{ragan2018}.

How to mask or partially mask an attribute depends on the attribute's type.
There are four types: fixed length string, variable length string, date, and categorical.
For example, for variable string attributes such as a person's name, the masked and partially masked value is calculated by Damerau-Levenshtein distance, which allows insertions, deletions, substitutions, and transposition.
For date attributes such as ``DoB'' (Date of Birth), \textit{masked} and \textit{partially disclosed} modes include inter and intra element discrepancies between year, month, and day.
Table \ref{tab:displaymode} shows some examples of partially disclosed value for different types of attributes.

At the beginning of the human interaction process, all the information is fully masked (i.e., legally de-identified mode), and the user can incrementally reveal the attributes as needed, increasing the level of disclosure for the record pair.




\section{Privacy Risk Score}
\label{sec:PrivacyScore}

In order to minimize the privacy risk associated with access to PII, the presented on-demand disclosure method requires a method for risk quantification. 
In this section, we define a quantitative measure, called the \emph{$k$-Anonymized Privacy Risk (KAPR) score}, for the identity disclosure risk associated with partial information disclosure.
We illustrate by example how KAPR is calculated, and we state and prove the desired properties of the measure. 


\subsection{Preliminaries}
The attributes occurring in a data table can be categorized into three non-disjoint sets:
\begin{enumerate}
\item \emph{identifiers} (e.g. name, social security number), which directly identify a person;
\item \emph{quasi-identifiers} (e.g. gender, ZIP code, race), which can be used to link records to external identifiers; and
\item \emph{sensitive attributes} (e.g. disease status, income), whose values for any individual must not be disclosed.
\end{enumerate}
A standard approach to preserving privacy is \emph{deidentification}, which refers to removing the identifiers from the data before sharing. This is not sufficient to guarantee anonymity, as quasi-identifiers can potentially be used to link records to identify records in other data sets, thereby  causing \emph{identity disclosure} of individuals within the data set. 


We use a privacy risk score based on \emph{$k$-anonymity}~\cite{Samarati2001, Sweeney2002} to measure the risk of identity disclosure. This property places a restriction on the disclosure of identifiers and quasi-identifiers.

\begin{definition} A data set is said to satisfy \emph{$k$-anonymity} for $k > 1$ if, for each combination of identifiers and quasi-identifiers, at least $k$ records exist in the data set sharing that combination. 
\end{definition}

No person in a $k$-anonymous data set can be distinguished from at least $k-1$ other people, which helps guard against identity disclosure (in other words, no particular person's data can be identified with certainty as belonging to them). $k$-anonymity alone is not sufficient to protect against attribute disclosure (e.g., cancer status)~\cite{DomingoFerrer2008}. However, when protection against attribute disclosure is guaranteed by separation of the sensitive attributes from the identifiers, it can be leveraged to create a useful measure for assessing the risk of identity disclosure (some amount of which is inherent in the record linkage context). 


\subsection{KAPR privacy risk score}

Before defining the KAPR score, we formalize the concept of a \emph{partial display}.
\begin{definition}[partial display]
Let $X$ be a data set with $N$ records and $D$ non-sensitive attributes, containing potential pairs of records to be linked. A partial display $\mathcal{X}$ is a set of $2n$ rows and $D$ attributes obtained from $n \leq \binom{N}{2}$ pairs of records from $X$ that are displayed to a user performing interactive record linkage. The rows in $\mathcal{X}$ are many-to-one related to the records in $X$. The values of the attributes in $\mathcal{X}$ may be fully disclosed, partially disclosed or masked versions of the corresponding attributes in $X$, as described in Table~\ref{tab:displaymode}.
\end{definition}

\begin{table*}[t]
  \centering
  \caption{Example of display mode for different attributes}
  \label{tab:displaymode}
    \begin{tabular}{lllll}
    \toprule
    Attribute&Type \hspace{0.5cm}  &Masked&Partial&Full\\
    \midrule
    \multirow{2}{2em}{ID} & \multirow{2}{2em}{String} & *********@ & *********9 & 1742682819\\
    & & *****\&**** & *****6**** & 1742668281\\
    \multirow{2}{2em}{Name} & \multirow{2}{2em}{Varchar} & *******@@@ & ******* JR & SANCHEZ JR\\
    & & ******* & ******* & SANCHEZ\\
    \multirow{2}{2em}{DOB} & \multirow{2}{2em}{Date} & @@/\&\&/**** & 08/09/**** & 08/09/1964\\
    & & \&\&/@@/**** & 09/08/**** & 09/08/1964\\
    \multirow{2}{2em}{Race} & \multirow{2}{2em}{Category} & @ & N/A & White\\
    & & \& & N/A & Asian\\
    \bottomrule
  \end{tabular}
\end{table*}

Underlying the partial displays is the closely related concept of an \emph{information disclosure state}, which in turn requires the concepts of \emph{proportion of characters disclosed} and \emph{anonymity set size}.

\begin{definition}[proportion of characters disclosed]
Let $\mathcal{X}$ be a partial display comprising $2n$ rows and $D$ attributes. Let the number of characters of attribute $j$ of record $i$ be $n_{ij}$. Let the number of characters whose true values are disclosed be $d_{ij}$. Then, the \textit{proportion of characters disclosed} for attribute $j$ of record $i$ is $p_{ij} \coloneqq d_{ij}/n_{ij}$.
\end{definition}

\begin{definition}[anonymity set size]
The \emph{anonymity set size} $k_i$ for a row $r_i$ in a partial display $\mathcal{X}$, based on an underlying data set $X$, is the number of records in $X$ that could correspond to $r_i$, based on the level of disclosure in $r_i$.
\end{definition}


\begin{definition}[information disclosure state]
Let $\mathcal{X}$ be a partial display comprising $2n$ rows and $D$ attributes. Let $p_{ij}$ represent the proportion of characters of attribute $j$ disclosed for row $i$. Let $k_i$ be the size of the anonymity set of record $i$ based on the information that has been disclosed. The \textit{information disclosure state} of $\mathcal{X}$ is defined by the matrix $\mathfrak{X} \in [0,1]^{2n \times D}$, with matrix elements $\mathfrak{X}_{ij} := (p_{ij}/k_i) \in [0,1]$. 
\end{definition}

There is a many-to-one mapping from partial displays to information disclosure states. We can now define KAPR as a function of an information disclosure state.

\begin{definition}[$k$-Anonymized Privacy Risk (KAPR) Score]
\label{kapr_def} Let $\mathfrak{X}$ be the information disclosure state associated with a partial display $\mathcal{X}$ with $2n$ records and $D$ attributes. Let $p_{ij}$ be the proportion of characters of attribute $j$ of record $i$ disclosed. Let $\kappa$ be the minimum allowed anonymity set size and let $k_i$ be the anonymity set size of record $i$ based on the current disclosure state. The \textit{k-Anonymized Privacy Risk score} (KAPR) is given by  
\begin{equation}
\label{kapr_def_eq}  K (\kappa; \mathfrak{X}) \coloneqq \frac{\kappa}{ND}\norm{\mathfrak{X}}_{1,1} = \frac{\kappa}{ND}\sum_{i=0}^{2n-1}\frac{1}{k_i}\sum_{j=0}^{D-1} \abs{p_{ij}}.
\end{equation}
Here, $\norm{\cdot}_{1,1}$ is the standard $L_{1,1}$ matrix norm.
\end{definition}

\subsection{Example}
\begin{table}[!htb]
	  \caption{Fully disclosed individual-level attributes. Colors indicate individuals (ground truth). There are three different individuals; rows 3 and 4 correspond to the same individual.}
	  \label{tab:microdata}
  \centering
  \begin{tabular}{lllll}
    \toprule
    $\texttt{ID}$ & $\texttt{Name}$ & $\texttt{DOB}$ &$\texttt{Race}$ & $\texttt{Income}$\\
    \midrule
    \rowcolor{mary1} {1} & {Mary} & {08/09/1964} & {Hispanic} & 69,426\\
    \rowcolor{mark2} {2} & {Mark} & {08/09/1964} & {Hispanic} & 38,001 \\
    \rowcolor{mary34} {3} & {Mary} & {09/08/1964} & {Black} & 27,998\\
    \rowcolor{mary34} {4} & {Mary} & {09/08/1964} & {Black} & 27,989\\
    \hline
    \bottomrule
  \end{tabular}
 \end{table}

\begin{table*}
  \centering
    \begin{minipage}{0.4\textwidth}
    \caption{Masked pairs.}
    \label{tab:masked}
    \begin{tabular}{lllllllll}
      \toprule
      $\texttt{ID}$ & $i$ & $\texttt{Name}$ & $\texttt{DOB}$ &$\texttt{Race}$& $k_i$ & ${\mathbf{p}_i}$ & $K_i$\\    
      \midrule
      \rowcolor{mary1} {(1,2)} & {1} & {***@} & {**/**/****} & {N/A} & 4 & (0, 0, 0) & 0\\
      \rowcolor{mark2} {} & {2} & {***\&} & {**/**/****} & {N/A} & 4 & (0, 0, 0) & 0\\
      \hline
      \rowcolor{mary1} {(1,3)} & {3} & {****} & {*@/*\&/****} & {N/A} & 4 & (0, 0, 0) & 0\\
      \rowcolor{mary34} {} & {4} & {****} & {*\&/*@/****} & {N/A} & 4 & (0, 0, 0) & 0\\
      \hline
      \rowcolor{mary1} {(1,4)} & {5} & {****} & {*@/*\&/****} & {N/A} & 4 & (0, 0, 0) & 0\\
      \rowcolor{mary34} {} & {6} & {****} & {*\&/*@/****} & {N/A} & 4 & (0, 0, 0) & 0\\
      \hline
      \rowcolor{mark2} {(2,3)} & {7} & {***@} & {*@/*\&/****} & {N/A} & 4 & (0, 0, 0) & 0\\
      \rowcolor{mary34} {} & {8} & {***\&} & {*\&/*@/****} & {N/A} & 4 & (0, 0, 0) & 0\\
      \hline
      \rowcolor{mark2} {(2,4)} & {9} & {***@} & {*@/*\&/****} & {N/A} & 4 & (0, 0, 0) & 0\\
      \rowcolor{mary34} {} & {10} & {***\&} & {*\&/*@/****} & {N/A} & 4 & (0, 0, 0) & 0\\
      \hline
      \rowcolor{mary34} {(3,4)} & {11} & {****} & {**/**/****} & {N/A} & 4 & $(0, 0, 0)$ & 0\\
      \rowcolor{mary34} {} & {12} & {****} & {**/**/****} & {N/A} & 4 & $(0, 0, 0)$ & 0\\
      \hline
      {} & {} & {} & {} & {} & {} & $K~(=\sum_i K_i) $ & $0$\\
      \bottomrule
    \end{tabular}
  \end{minipage}
  
  \begin{minipage}{0.4\textwidth}
  	\caption{Partial disclosure.}
	\label{tab:partial}
    \begin{tabular}{llllllll}
      \toprule
      $\texttt{ID}$ & $i$ & $\texttt{Name}$ & $\texttt{DOB}$ &$\texttt{Race}$& $k_i$ & ${\mathbf{p}_i}$ & $K_i$\\    
      \midrule
      \rowcolor{mary1} {(1,2)} & {1} & {***y} & {**/**/****} & {N/A} & 3 & (1/4, 0, 0) & 1/432\\
      \rowcolor{mark2} {} & {2} & {***k} & {**/**/****} & {N/A} & 1 & (1/4, 0, 0) & 1/144\\
      \hline
      \rowcolor{mary1} {(1,3)} & {3} & {****} & {*8/*9/****} & {N/A} & 1 & (0, 2/8, 0) & 1/144\\
      \rowcolor{mary34} {} & {4} & {****} & {*9/*8/****} & {N/A} & 2 & (0, 2/8, 0) & 1/288\\
      \hline
      \rowcolor{mary1} {(1,4)} & {5} & {****} & {*8/*9/****} & {N/A} & 1 & (0, 2/8, 0) & 1/144\\
      \rowcolor{mary34} {} & {6} & {****} & {*9/*8/****} & {N/A} & 2 & (0, 2/8, 0) & 1/288\\
      \hline
      \rowcolor{mark2} {(2,3)} & {7} & {***k} & {*8/*9/****} & {N/A} & 1 & (1/4, 2/8, 0) & 1/72\\
      \rowcolor{mary34} {} & {8} & {***y} & {*9/*8/****} & {N/A} & 2 & (1/4, 2/8, 0) & 1/144\\
      \hline
      \rowcolor{mark2} {(2,4)} & {9} & {***k} & {*8/*9/****} & {N/A} & 1 & (1/4, 2/8, 0) & 1/72\\
      \rowcolor{mary34} {} & {10} & {***y} & {*9/*8/****} & {N/A} & 2 & (1/4, 2/8, 0) & 1/144\\
      \hline
      \rowcolor{mary34} {(3,4)} & {11} & {****} & {**/**/****} & {N/A} & 3 & $(0, 0, 0)$ & 0\\
      \rowcolor{mary34} {} & {12} & {****} & {**/**/****} & {N/A} & 3 & $(0,0,0)$ & 0\\
      \hline
      {} & {} & {} & {} & {} & {} & $K~(=\sum_i K_i) $ & $0.072$\\
      \bottomrule
    \end{tabular}
  \end{minipage}

      \small
	  \caption{Full disclosure.}
	  \label{tab:full}
  \begin{minipage}{0.4\textwidth}
  \begin{tabular}{llllllll}
    \toprule
    {\texttt{ID}} & $i$ & $\texttt{Name}$ & $\texttt{DOB}$ & $\texttt{Race}$ & $k_i$ & $\mathbf{p}_i$ & $K_i$\\
    \midrule
    \rowcolor{mary1} {(1,2)} & 1 & {Mary} & {08/09/1964} & {Hispanic} & 1 & $(1,1,1)$ & 1/12\\
    \rowcolor{mark2} {} & 2 & {Mark} & {08/09/1964} & {Hispanic} & 1 & $(1,1,1)$ & 1/12\\
    \hline
    \rowcolor{mary1} {(1,3)} & 3 & {Mary} & {08/09/1964} & {Hispanic} & 1 & $(1,1,1)$ & 1/12\\
    \rowcolor{mary34} {} & 4 & {Mary} & {09/08/1964} & {Black} & 2 & $(1,1,1)$ & 1/24\\
    \hline
    \rowcolor{mary1} {(1,4)} & 5 & {Mary} & {08/09/1964} & {Hispanic}  & 1 & $(1,1,1)$ & 1/12\\
	\rowcolor{mary34} {} & 6 & {Mary} & {09/08/1964} & {Black} & 2 & $(1,1,1)$ & 1/24\\
    \hline
    \rowcolor{mark2} {(2,3)} & 7 & {Mark} & {08/09/1964} & {Hispanic} & 1 & $(1,1,1)$ & 1/12\\
    \rowcolor{mary34} {} & 8 & {Mary} & {09/08/1964} & {Black} & 2 & $(1,1,1)$ & 1/24\\
    \hline
    \rowcolor{mark2} {(2,4)} & 9 & {Mark} & {08/09/1964} & {Hispanic} & 1 & $(1,1,1)$ & 1/12\\
	\rowcolor{mary34} {} & 10 &  {Mary} & {09/08/1964} & {Black} & 2 & $(1,1,1)$ & 1/24\\
    \hline
    \rowcolor{mary34} {(3,4)} & 11 & {Mary} & {09/08/1964} & {Black} & 2 & $(1,1,1)$ & 1/24\\
	\rowcolor{mary34} {} & 12 & {Mary} & {09/08/1964} & {Black} & 2 & $(1,1,1)$ & 1/24\\
	\hline
    {} & {} & {} & {} & {} & {} & $K~(=\sum_i K_i) $ & 0.750\\
    \bottomrule
  \end{tabular}
  \end{minipage}
\end{table*}
We illustrate the utility of KAPR via an example. Consider Table~\ref{tab:microdata}, containing microdata about individual people, whose schema comprises the identifier $\left\lbrace\texttt{Name}\right\rbrace$, the quasi-identifiers $\left\lbrace\texttt{DOB},~\texttt{Race}\right\rbrace$, and the sensitive attribute $\left\lbrace\texttt{Income}\right\rbrace$. The data linkage task to be accomplished is to merge duplicate records in this data set. This is accomplished by having a worker manually inspect all pairs of similar records and decide whether or not they correspond to the same individual. In this example, the ground truth is that records 1 and 2 belong to unique individuals, while records 3 and 4 correspond to the same individual (as indicated by the row colors). Because all of the records in this table are quite similar, the user interface will display every possible pair of records from this table to the worker.

Initially, the pairs will be displayed in masked mode. This corresponds to the information disclosure state depicted in Table~\ref{tab:masked}. The field \texttt{ID} contains a tuple denoting which records from the base data set (Table~\ref{tab:microdata}) constitute this pair, and the field $i$ is used to label the rows of data displayed. Since \texttt{Income} is a sensitive attribute, it will not be displayed. We denote by $\mathbf{p}_i$ the vector whose entries consist of the proportions of characters of the \texttt{Name}, \texttt{DOB} and \texttt{Race} fields respectively. We denote by $k_i$ the size of the anonymity set for record $i$. In principle, any number of the characters of \texttt{Name} and \texttt{DOB} can be disclosed. Because \texttt{Race} is a Category variable, it can only be not disclosed or fully disclosed; there is no option for partial disclosure. In this mode, each row $i$ in the display has an anonymity set size of $k_i = 4$, because the masked data could correspond to any of the records in Table~\ref{tab:microdata}. However, no characters of any of the three attributes are disclosed, meaning $\mathbf{p}_i = (0,0,0)$ for each row $i$, and hence $K_i = 0$ for each row as well. Thus, the KAPR score for this information disclosure state is 0.

The incremental display allows the user to reveal incremental additional information about any cell in the display, and the privacy cost of this action is quantified by the amount by which the KAPR score is incremented as result. As an example, the user may have decided to switch from masked mode to partial mode for every cell in the display, as depicted in Table~\ref{tab:partial}. This will change the values of $k_i$ and $\mathbf{p}_i$ for each row. For example, for row $i=1$, the size of the anonymity set $k_i$ has decreased from 4 in masked mode to 3 in partial mode, since we now know that the last character of \texttt{Name} is `y', which means it must correspond to one of the three rows where \texttt{Name} = `Mary' in Table~\ref{tab:microdata}. Furthermore, we have now revealed $1/4$ of the characters of \texttt{Name}, but still no characters of the other attributes, meaning $\mathbf{p}_i = (1/4,0,0)$ now. In this case, since $N=12$, $D=3$, this means that $K_i = 1/432$. The value of $K_i$ for each row is shown separately; for $i \in \lbrace 11,12 \rbrace$, $K_i = 0$ still, because the identifying attributes of the underlying records are identical, and thus no characters need be revealed. The KAPR score for this information disclosure state is $\sum_{i=1}^{12} K_i = 31/432 \approx 0.072 > 0$. So the KAPR score has increased incrementally from $0$ when every row was displayed in masked mode. It should be clear from the example that the KAPR score for this information disclosure state does not depend upon the order in which the information was disclosed; regardless of the order in which the user clicked the cells of the display to reveal more information, the final information disclosure state is the same, and the KAPR score depends only on the information disclosure state.

The maximum amount of information that can be revealed is for every row to be displayed in full disclosure mode. This information disclosure state is depicted in Table~\ref{tab:full}. In this state, $\mathbf{p}_i = (1,1,1)$ for every row $i$. Even in this fully disclosed state, $k_i = 2$ for some rows (the ones where the underlying record is one of the two duplicates in the original record set with respect to the identifying attributes). As a result, the KAPR score is 0.750, which is less than the theoretical maximum value of 1 for a data set without duplicates.

\subsection{Properties of KAPR score}

KAPR has several appealing and easily demonstrated properties that make it a useful measure of identity disclosure risk for interactive record linkage. 

\begin{property} The KAPR score $K: \mathbb{R}^{N \times D} \rightarrow \mathbb{R}_+$ is a norm (it is non-negative, homogeneous, and obeys the triangle inequality). Hence, it can be assigned the geometrical interpretation of the ``length'' of the information disclosure state, where ``longer'' means ``higher risk of identity disclosure''.
\end{property}

\begin{proof}
This follows directly from Definition~\ref{kapr_def} and the fact that $\kappa/ND > 0$, since a norm multiplied by a positive constant is still a norm (it represents a uniform rescaling of all lengths).
\end{proof}

\begin{property} KAPR score manifestly depends only on the information disclosure state, which in turn depends only on the information displayed to the user, and not on the precise sequence of actions taken by the user to lead to that state.
\end{property}

\begin{property}
KAPR score ranges from 0 (when no characters are disclosed, such as when every attribute is in masked display mode) to 1 (when every character is disclosed and the anonymity set size of every record in the underlying data set is 1).
\end{property} 

\begin{proof}
$K \geq 0$ follows from KAPR being a norm. $K \leq 1$ follows from the fact that $k_i \geq \kappa, \abs{p_{ij}} \leq 1 \Rightarrow \abs{p_{ij}}/k_i \leq \sum_{i=0}^{N-1} \sum_{j=0}^{D-1} 1/\kappa =ND/\kappa \Rightarrow K(\kappa; X) =(\kappa/ND)\norm{X}_{1,1} \leq 1.$
\end{proof}

\begin{property} \label{kapr_inc} KAPR score monotonically increases as more characters are disclosed.
\end{property}
\begin{proof}
Consider the partial derivative of $K(\kappa, X)$ with respect to any of the $p_{ij}$: $\frac{\partial K}{\partial p_{ij}} = \frac{\kappa}{ND} \sum_{r,s} \frac{1}{k_r} \frac{\partial \abs{p_{ij}}}{\partial p_{rs}} = \frac{\kappa}{ND} \sum_{r,s} \frac{1}{k_r} \delta_{ir} \delta_{js} = \frac{\kappa}{ND} \frac{1}{k_i} > 0$.
Thus, KAPR always increases with disclosure of new characters.
\end{proof}

\begin{property}
The KAPR score penalty for disclosing attribute values for any particular record is a monotonically decreasing function of the anonymity set size for the record. 
\end{property}
\begin{proof}
Suppose characters from attribute $j$ of record $i$ are disclosed. From the proof of Property~\ref{kapr_inc} above, $\frac{\partial K}{\partial p_{ij}}=\frac{\kappa}{ND} \frac{1}{k_i}$. This is a decreasing function of $k_i$, the anonymity set size of record $i$. Moreover, disclosure can only decrease or maintain the anonymity set size, so a given disclosure that causes a reduction in the anonymity set size causes a greater increase in KAPR than the same disclosure would if the anonymity set size was preserved.
\end{proof}

\section{Conclusion}
\label{sec:Conclusion}
Integrating data records from multiple databases is crucial for many applications and research. Interactive record linkage is often used to ensure high linking quality. Therefore, there has been increasing concerns about privacy on interactive record linkage. We proposed a measurement called KAPR to quantify the privacy risk for interactive record linkage projects, and showed how it meet the desired properties. By using KAPR, data custodians and researchers can better communicate and agree on the optimal trade-off point for legitimate access to data and privacy protection.

\bibliographystyle{plain}
\bibliography{sample-bibliography}

\end{document}